\documentclass[fleqn,usenatbib]{mnras}

\usepackage{newtxtext,newtxmath}

\usepackage[T1]{fontenc}

\DeclareRobustCommand{\VAN}[3]{#2}
\let\VANthebibliography\thebibliography
\def\thebibliography{\DeclareRobustCommand{\VAN}[3]{##3}\VANthebibliography}

\usepackage{graphicx}	
\usepackage{amsmath}	
\usepackage{lscape}

\newcommand{\amend}{}

\title[Searching for Outbursts in 67P Photometry]{Searching for Outbursts in the Ground-Based Photometry of 67P/Churyumov-Gerasimenko}

\author[D. Gardener et al.]{
Daniel Gardener,$^{1}$\thanks{E-mail: dgar@roe.ac.uk}
Colin Snodgrass,$^{1}$
Nicolas Ligier$^{2,3}$
\\
$^{1}$Institute for Astronomy, University of Edinburgh, Blackford Hill, Edinburgh EH9 3HJ, UK\\
$^{2}$Institut d’Astrophysique Spatiale, CNRS/Université Paris-Saclay, B'atiment 121, F-91405 Orsay, France\\
$^{3}$School of Physical Sciences, The Open University, Milton Keynes MK7 6AA, UK\\
}

\date{Accepted XXX. Received YYY; in original form ZZZ}

\pubyear{2022}

\begin{document}
\label{firstpage}
\pagerange{\pageref{firstpage}--\pageref{lastpage}}
\maketitle

\begin{abstract}
67P/Churyumov-Gerasimenko is a Jupiter-family comet that was the target of the Rosetta mission, the first mission to successfully orbit and land a probe on a comet. This mission was accompanied by a large ground-based observing campaign. We have developed a pipeline to calibrate and measure photometry of comet 67P during its 2016 perihelion passage, making use of all visible wavelength broadband imaging collected across a wide range of facilities. The pipeline calibrates the brightness of the comet to a common photometric system (Pan-STARRS 1) using background stars within the field allowing for compilation and comparison of multiple data sets. Results follow the predictions based on previous apparitions: 67P shows no obvious change in activity levels from orbit-to-orbit and coma colours remain constant throughout the apparition. We detected an outburst on 2015 August 22 of $\sim$0.14 mag. The brightness and estimated mass of this outburst puts it in line with the outbursts directly observed on the nucleus by Rosetta. An in situ outburst was observed at the same time as the one seen from the ground, however linking these two events directly remains challenging. 
\end{abstract}




\section{Introduction}

67P/Churyumov--Gerasimenko was the target of the Rosetta mission, the first mission to successfully orbit a cometary nucleus and follow it along its journey through perihelion. The mission returned a unique cache of data, collected in situ at the nucleus, revealing new insights about comet surface activity \citep[e.g.][]{El-Maarry2019, Filacchione2019, Vincent2019, Marschall2020, Choukroun2020, Mottola2020}. This mission was backed up by a large ground-based observing campaign \citep{Snodgrass2017} that followed the activity of 67P through its perihelion passage. This data set is one of the most detailed and comprehensive data sets ever taken of a comet, with coverage across almost all of the comet's inward and outward journeys, so provides an ideal treasure trove for detailed analysis. 

The Rosetta mission provides us with an opportunity to link ground-based observations with events observed in situ in orbit around the comet's nucleus. Outbursts are a signature of activity; many were observed on the nuclear surface by instruments onboard Rosetta. Inbound to the comet an outburst was detected in 2014 April \citep{Tubiana2015}. The comet was regularly monitored as the spacecraft approached between 2014 July and 2014 October, with no further outbursts seen. Once in orbit around the comet an outburst was seen in 2015 February \citep{Knollenberg2016}. Over the next few months, the Rosetta probe had to retreat to a safe distance from the comet due to high dust content in the coma; during this time any outbursts on the comet's surface could have been missed. Between July and September 2015, as the comet passed perihelion, 34 individual outbursts were observed as detailed in \citet{Vincent2016}. At the same time, \citet{Boehnhardt2016} saw a large dust ejection event \amend{in the coma morphology in images acquired at} the \amend{2-m} Wendelstein telescope on 2015 August 22--23, but they do not make a link between this observation and any outbursts seen by Rosetta. \amend{\citet{Knight2017}, observing from the 0.8-m Lowell telescope, also saw the same outburst in their photometry on 2015 August 22. They make a tentative link to an outburst observed by Rosetta. They also report a possible outburst occurring on 2015 September 19 but they do not match it with any other known outbursts of 67P.} Another notable outburst was seen by multiple instruments on Rosetta on 2016 February 19 \citep{Grun2016}. Initial analysis of TRAPPIST observations over this period by \citet{Grun2016} claim to show an increased and sustained brightness correlating to this outburst. \citet{Agarwal2017} saw an outburst on 2016 July 3.

Aside from searching for small-scale transient events, tracing activity can give us an insight into the ageing processes that affects a comet. Predictions of the dust activity were made by \citet{Snodgrass2013} and the observations have shown the comet to be following these predictions \citep{Snodgrass2017}. This leads us to believe that the activity of 67P remains largely unchanged from orbit-to-orbit and therefore results from Rosetta can be applied more generally to help constrain models of comet activity evolution and scale results to different comets and apparitions. The activity analysis performed in \citet{Snodgrass2017} was made using an approximate calibration; in this paper we detail a precise calibration method using comparison to the Pan-STARRS catalogue \citep{Tonry2012}. The calibration method was applied across the majority of the \citet{Snodgrass2017} campaign data. In this paper we search the broadband photometry to find small-scale variations that could be linked to outbursts. The large pool of data allows us to confirm the brightening across multiple data sets, paying attention to events seen by Rosetta to see if any links could be made. We also look to see if we can confirm the outbursts seen in \citet{Vincent2016}, \citet{Boehnhardt2016}, \amend{\citet{Knight2017},} \citet{Grun2016} and \citet{Agarwal2017}. We aim to constrain the detectability of small-scale events from ground-based observations. These constraints will help future interpretations of ground-based observations of comets and link them to changes in the nucleus, which we cannot visit directly. 

\section{Observations}

\begin{table}
    \centering
    \caption{Summary table of analysed observations. Filters in letters for standard bands, with lowercase (\textit{g,r,i,z}) indicating Sloan Digital Sky Survey (SDSS)-type filters and upper case (\textit{B,V,R,I}) indicating Johnson/Cousins types.}
    \label{tab:Observations}
    \begin{tabular}{llll}
        \hline
        Telescope/Instrument               & filter           & dates (YY/MM/DD)  \\
        \hline
        NOT/ALFOSC                         & \textit{V,R}     & 13/05/13-16/08/10 \\
        NOT/StanCam                        & \textit{V,R}     & 14/04/05-16/05/22 \\
        OGS/SDC                            & visible          & 14/09/21-16/07/04 \\
        TRAPPIST-South/CCD                 & \textit{B,V,R,I} & 15/04/18-16/06/07 \\
        NTT/EFOSC                          & \textit{r}       & 15/04/22-16/07/29 \\
        VLT/FORS                           & \textit{R}       & 15/05/21-17/03/25 \\
        WHT/ACAM                           & \textit{R,I}     & 15/07/06-16/06/28 \\
        STELLA/WIFSIP                      & \textit{g,r,i,z} & 15/07/18-16/06/08 \\
        LT/IO:O                            & \textit{g,r,i,z} & 15/07/18-16/06/11 \\
        LOT                                & \textit{B,V,R}   & 15/08/02-15/11/07 \\
        LCOGT/Merope                       & \textit{g,r,i,z} & 15/08/07-15/09/21 \\
        Rozhen BNAO 2-m/FoReRo-2           & \textit{B,R}     & 15/08/11-16/11/06 \\
        CA 2.2-m/CAFOS                     & \textit{R}       & 15/08/14-16/06/05 \\
        CA 3.5-m/MOSCA                     & \textit{R}       & 15/08/18-15/08/25 \\
        Lowell 0.8-m/NASAcam               & \textit{R}       & 15/08/18-15/12/01 \\
        TNG/DOLoRes                        & \textit{B,V,R}   & 15/08/18-16/03/17 \\
        Wendelstein 2-m/WWFI               & \textit{g,r,i}   & 15/08/21-16/05/08 \\
        OSN 1.5-m/CCD                      & \textit{R}       & 15/09/21-16/04/30 \\
        INT/WFC                            & \textit{B,r,i}   & 15/10/13-16/06/23 \\
        BTA/SCORPIO2                       & \textit{r,g}     & 15/11/07-16/04/01 \\
        LCOGT/SBIG                         & \textit{r}       & 15/12/14-16/01/30 \\
        OSN 0.9-m/CCD                      & \textit{R}       & 16/01/13-16/01/16 \\
        LCOGT/Sinistro                     & \textit{r}       & 16/01/27-16/03/30 \\
        Gemini N/GMOS                      & \textit{g,r,i,z} & 16/02/16-16/05/28 \\
        IRTF/MORIS                         & \textit{r}       & 16/03/14-16/03/28 \\
        \hline
        \amend{LT/IO:O}                    & \amend{\textit{g,r,i,z}} & \amend{21/07/06-22/06/11} \\
        \hline
    \end{tabular}
\end{table}

Table~\ref{tab:Observations} summarises the broadband imaging observations of 67P made between 2013 and 2017. The whole data set covers a total on-target observing time of $\sim$640 hrs with 9606 individual frames from 27 telescopes across 9 countries. The data at the beginning and end of the campaign offered limited use due to our pipeline's limitation in detecting dim targets in crowded fields and therefore are not suitable for automatic photometry analysis. Nordic Optical Telescope (NOT) observations, for example, cover a period in 2014 when the comet was visible at low altitudes in the northern sky. These data however, due to the comet's faintness and the high airmass during observation, are not suitable for automatic processing. Analysis of the 2014 NOT data was performed by \citet{Zaprudin2015} and analysis of the remaining 2014 data can be found in \citet{Snodgrass2016a}. In this paper we focus our analyses on the data between 2015--2016, which covers the majority of the comet's inner Solar System passage and the 'escort' phase of the Rosetta mission. It was during this phase that the majority of the telescopes were observing the comet regularly, providing almost 24/7 coverage at some points. The early portion of this data were afflicted by less favourable viewing conditions due to the small solar elongation in the early months of observation, including during the perihelion on 2015 August 13. After 2015 October it became more favourable to view.

We highlight data that have unique coverage or significance in the following sub-sections:

\subsection{VLT}

The 8-m European Southern Observatory Very Large Telescope (VLT) in Chile provides the longest observing period from start to end; the VLT began observing 67P in 2013 to measure the astrometry of the comet before the spacecraft's arrival and constrain the start of activity \citep{Snodgrass2016a}. It also extends beyond the observing windows of most other telescopes with observations made until 2017 March 25 providing unique coverage of the comet's outward journey as it dims to below detectable brightness. 

\subsection{NOT}

Similarly to the VLT, the NOT started observing 67P in 2013. Despite being located on La Palma in the northern hemisphere, the 2.56-m telescope is capable of observing at very low altitudes meaning it started observing before its northern hemisphere counterparts. However, these early observations are of limited use because of the difficulty of detecting the faint comet. The NOT provides regular coverage over the course of the perihelion passage, observing once or twice a week between 2015 July 1 and 2016 August 10 in both \textit{R}- and \textit{V}-bands. More details of these observing runs can be found in \citet{Zaprudin2015} and \citet{Zaprudin2017}.

\subsection{TRAPPIST}

The robotic 60-cm TRAPPIST telescope in La Silla \citep{Jehin2011} provides regular coverage across all of the perihelion passage including a unique period between 2015 April 18 and June 27 when the comet was difficult to observe from northern hemisphere observatories. TRAPPIST provided observations in \textit{B}-, \textit{V}-, \textit{R}- and \textit{I}-bands across the whole passage. For more details see \cite{Snodgrass2016b}.

\subsection{LT}

The robotic 2-m Liverpool Telescope (LT) on La Palma provides some of the most regular coverage in \textit{r}-band across the majority of the perihelion passage between 2015 July 18 and 2016 June 11 and measurements in the \textit{g}-, \textit{i}- and \textit{z}-bands between 2015 July 18 to August 31 and 2016 February 10 to June 11. This run was also detailed in \cite{Snodgrass2016b}.

\amend{Using the LT, we undertook regular monitoring of 67P during its next apparition between 2021 July 6 and 2022 June 11 covering both inbound and outbound journeys. These observations are discussed in Section~\ref{sec:2021}.}

\subsection{Wendelstein}

The 2-m telescope at the Wendelstein observatory in Germany provided over 90 hours of regular post-perihelion coverage between 2015 August 22 and 2016 May 9 and shows initial evidence for an outburst \citep{Boehnhardt2016}.  

\subsection{Lowell}
\amend{The 0.8-m telescope at the Lowell Observatory made regular observations post-perihelion between 2015 August 18 and 2015 December 1. It observed the same outburst seen by the Wendelstein telescope as well as a second potential outburst \citep{Knight2017}.}

\section{Data reduction}

\begin{figure}
	\includegraphics[width=\columnwidth]{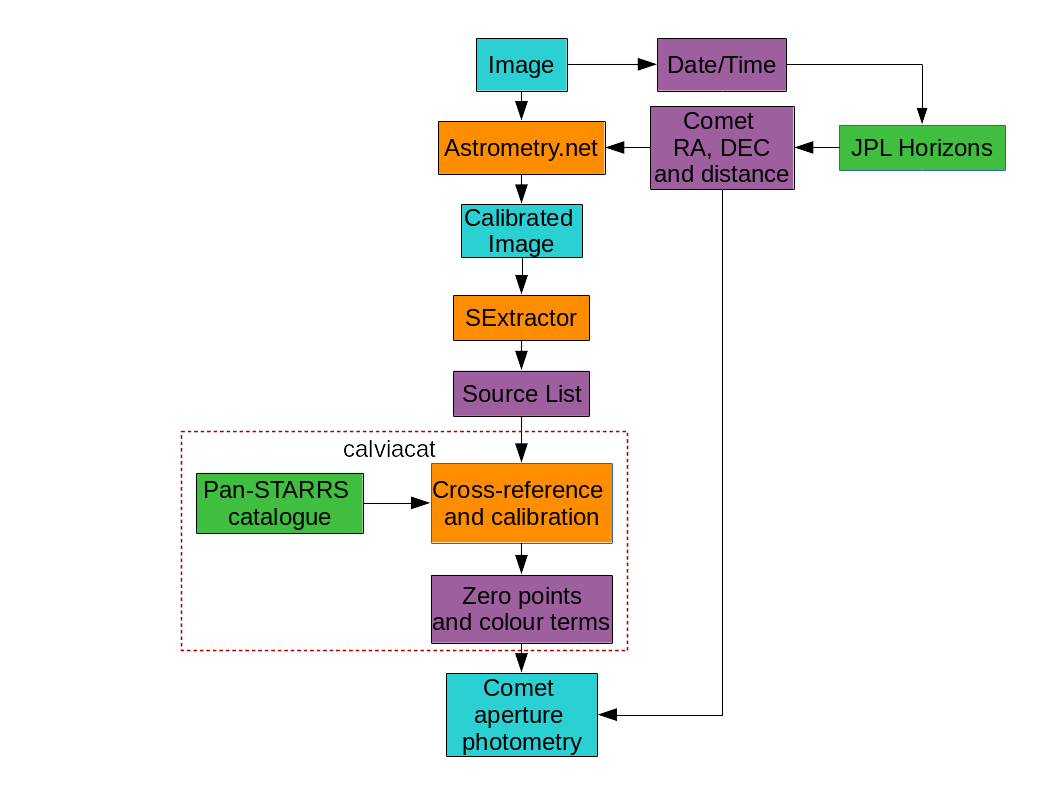}
    \caption{Flowchart showing the steps in the automatic astrometry and photometry calibration pipeline.}
    \label{fig:flowchart}
\end{figure}

Our data are calibrated through a custom-built pipeline, which incorporates JPL Horizons \citep{Giorgini1996}, Astrometry.net \citep{Lang2010}, \textsc{SExtractor} \citep{Bertin1996} and \textsc{calviacat} \citep{Kelley2019}. Figure~\ref{fig:flowchart} shows a flowchart of the steps in the pipeline. The steps in the pipeline are:
\begin{enumerate}
    \item The date from the FITS header is extracted and passed to JPL Horizons to retrieve the ephemeris data of the comet (for our purposes we are interested in position and geocentric distance). The RA and DEC of the target at that date is passed to Astrometry.net.
    \item Astrometry.net calibrates the image's world coordinate system (WCS) using the image coordinates of the sources in the field and cross-referencing them with its own catalogue to accurately determine the astrometry of the image. Astrometry.net searches within 1 degree of the comet coordinates passed from JPL Horizons to decrease the computation time compared with performing a blind search. 
    \item \textsc{SExtractor} extracts the instrumental magnitudes of all sources within the field of view using automatic elliptical apertures defined so as to contain at least 90 per cent of the source flux around every detected object \citep{Bertin1996}. \textsc{SExtractor} flags sources that could be problematic during the extraction process; these warnings can indicate neighbouring sources, saturated pixels or memory overflows. The flagged sources are removed from the source list before using the WCS to create a catalogue and passing it to \textsc{calviacat}. 
    \item The cleaned source list is then fed into \textsc{calviacat} which calibrates the magnitudes to the Pan-STARRS photometric system \citep{Tonry2012}. It works by cross-referencing the source list with the Pan-STARRS DR1 (PS1) catalogue. Using the WCS coordinates of sources within the frame it finds matching PS1 sources and then estimates the calibration constant with colour correction. Table~\ref{tab:Colour_terms} displays the colour terms used in the calibrations for each filter in each data set. Before calibrating the magnitudes, the pipeline removes any sources with a PS1 PSF - Kron magnitude greater than 0.05. These sources are likely to be galaxies and therefore cannot be used as calibrators.
    \item The comet is identified by finding the source with the WCS coordinates that most closely match the coordinates from JPL Horizons, within 4 arcsec. Photometry is then measured with a fixed aperture of 10,000 km radius using the geocentric distance from JPL Horizons and appropriate pixel scale from the WCS to calculate the corresponding radius in pixels. If other catalogue sources are found within the aperture, the pipeline raises a flag and records their Pan-STARRS magnitudes. 
\end{enumerate}

The pipeline does have limitations. In order to perform automatic astrometry and magnitude calibration, the pipeline needs a large sample of background stars which can be lacking in some frames, particularly from instruments with relatively small fields-of-view. The pipeline does not take into account bright field stars that can contribute to the background level inside the aperture, or completely outshine the comet, even if their centres are outside the comet aperture. This leads to a target being artificially brightened, especially when the target is dimmer and more likely to be outshone by field stars. These have to be removed manually. 

\begin{table}
    \centering
    \caption{Summary table of the colour terms used during colour calibration for each data set.}
    \label{tab:Colour_terms}
    \begin{tabular}{llll}
        \hline
        Telescope/Instrument      & filter     & colour        & colour term \\
        \hline
        NOT/ALFOSC                & \textit{V} & \textit{g--r} & 0.45 \\
                                  & \textit{R} & \textit{g--r} & 0.14\\
        OGS/SDC                   & visible    & \textit{g--r} & -0.41\\
        TRAPPIST-South/CCD        & \textit{B} & \textit{g--r} & -0.55\\
                                  & \textit{V} & \textit{g--r} & 0.48\\
                                  & \textit{R} & \textit{g--r} & 0.14\\
                                  & \textit{I} & \textit{r--i} & 0.23\\
        VLT/FORS                  & \textit{R} & \textit{g--r} & 0.22\\
        WHT/ACAM                  & \textit{r} & \textit{g--r} & -0.04\\
                                  & \textit{i} & \textit{r--i} & 0.04\\    
        STELLA/WIFSIP             & \textit{g} & \textit{g--r} & -0.15\\
                                  & \textit{r} & \textit{g--r} & -0.02\\
                                  & \textit{i} & \textit{r--i} & 0.07\\
                                  & \textit{z} & \textit{i--z} & -0.20\\
        LT/IO:O                   & \textit{g} & \textit{g--r} & -0.01\\
                                  & \textit{r} & \textit{g--r} & -0.02\\
                                  & \textit{i} & \textit{r--i} & 0.04\\
                                  & \textit{z} & \textit{i--z} & 0.13\\
        LOT                       & \textit{B} & \textit{g--r} & -0.28\\
                                  & \textit{V} & \textit{g--r} & 0.43\\
                                  & \textit{R} & \textit{g--r} & 0.16\\
        LCOGT/Merope              & \textit{g} & \textit{g--r} & 0.03\\
                                  & \textit{r} & \textit{g--r} & 0.01\\
                                  & \textit{i} & \textit{r--i} & 0.04\\
                                  & \textit{z} & \textit{i--z} & -0.07\\
        Rozhen BNAO 2-m/FoReRo-2  & \textit{R} & \textit{g--r} & 0.21\\
        CA 2.2-m/CAFOS            & \textit{R} & \textit{g--r} & 0.22\\
        CA 3.5-m/MOSCA            & \textit{R} & \textit{g--r} & 0.10\\
        Lowell 0.8-m/NASAcam      & \textit{R} & \textit{g--r} & 0.08\\
        TNG/DOLoRes               & \textit{B} & \textit{g--r} & -0.52\\
                                  & \textit{V} & \textit{g--r} & 0.42\\
                                  & \textit{R} & \textit{g--r} & 0.14\\
        Wendelstein 2-m/WWFI      & \textit{g} & \textit{g--r} & -0.02\\
                                  & \textit{r} & \textit{g--r} & 0.02\\ 
                                  & \textit{i} & \textit{r--i} & 0.05\\
        OSN 1.5-m/CCD             & \textit{R} & \textit{g--r} & 0.20\\
        INT/WFC                   & \textit{B} & \textit{g--r} & -0.45\\
                                  & \textit{r} & \textit{g--r} & 0.04\\
                                  & \textit{i} & \textit{r--i} & 0.08\\
        BTA/SCORPIO2              & \textit{r} & \textit{g--r} & 0.01\\
        LCOGT/SBIG                & \textit{r} & \textit{g--r} & 0.01\\
        OSN 0.9-m/CCD             & \textit{R} & \textit{g--r} & 0.09\\
        LCOGT/Sinistro            & \textit{r} & \textit{g--r} & 0.01\\
        Gemini N/GMOS             & \textit{g} & \textit{g--r} & -0.08\\
                                  & \textit{r} & \textit{g--r} & 0.10\\
                                  & \textit{i} & \textit{r--i} & 0.15\\
                                  & \textit{z} & \textit{i--z} & -0.27\\
        \hline
    \end{tabular}
\end{table}

\section{Results}

\begin{table*}
    \caption{Summary table of data processed through the pipeline. Frames input is the original number of images passed to the pipeline. Frames processed is the number of frames successfully calibrated by the pipeline, images causing the pipeline to fail or images manually removed from the final data are not included in this number.}
    \label{tab:Results}
    \begin{tabular}{lllllllllllllllll}
        \hline
        Telescope/instrument      & \multicolumn{16}{c}{Frames input / successfully processed} \\
                                  &\multicolumn{2}{c}{\textit{B}} & \multicolumn{2}{c}{\textit{V}} & \multicolumn{2}{c}{\textit{R}} & \multicolumn{2}{c}{\textit{I}} & \multicolumn{2}{c}{\textit{g}} & \multicolumn{2}{c}{\textit{r}} & \multicolumn{2}{c}{\textit{i}} & \multicolumn{2}{c}{\textit{z}}\\
        \hline
        NOT/ALFOSC                & --       & --        & 489       & 460       & 667       & 459       & --        & --        & --        & --        & --        & --        & --        & --        & --        & --   \\
        NOT/StanCam               & --       & --        & 51        & 0         &  56       & 0         & --        & --        & --        & --        & --        & --        & --        & --        & --        & --   \\
        OGS/SDC                   & --       & --        & --        & --        &  258      & 192       & --        & --        & --        & --        & --        & --        & --        & --        & --        & --   \\
        TRAPPIST-South/CCD        & 63       & 59        & 247       & 217       & 74        & 72        & 69        & 61        & --        & --        & --        & --        & --        & --        & --        & --   \\
        NTT/EFOSC                 & --       & --        & --        & --        & --        & --        & --        & --        & --        & --        & 24        & 0         & --        & --        & --        & --   \\
        VLT/FORS                  & --       & --        & --        & --        & 53        & 52        & --        & --        & --        & --        & --        & --        & --        & --        & --        & --   \\
        WHT/ACAM                  & --       & --        & --        & --        & --        & --        & --        & --        & --        & --        & 9         & 9         & 3         & 3         & --        & --   \\
        STELLA/WIFSIP             & --       & --        & --        & --        & --        & --        & --        & --        & 25        & 25        & 745       & 645       & 25        & 25        & 35        & 35   \\
        LT/IO:O                   & --       & --        & --        & --        & --        & --        & --        & --        & 109       & 100       & 355       & 317       & 109       & 100       & 109       & 100  \\
        LOT                       & 4        & 3         & 5         & 5         & 14        &  13       & --        & --        & --        & --        & --        & --        & --        & --        & --        & --   \\
        LCOGT/Merope              & --       & --        & --        & --        & --        & --        & --        & --        & 28        & 28        & 32        & 32        & 14        & 14        & 14        & 14   \\
        Rozhen BNAO 2-m/FoReRo-2  & 5        & 0         & --        & --        & 13        & 2         & --        & --        & --        & --        & --        & --        & --        & --        & --        & --   \\
        CA 2.2-m/CAFOS            & --       & --        & --        & --        & 912       & 690       & --        & --        & --        & --        & --        & --        & --        & --        & --        & --   \\
        CA 3.5-m/MOSCA            & --       & --        & --        & --        & 22        & 22        & --        & --        & --        & --        & --        & --        & --        & --        & --        & --   \\
        Lowell 0.8-m/NASAcam      & --       & --        & --        & --        & 354       & 318       & --        & --        & --        & --        & --        & --        & --        & --        & --        & --   \\
        TNG/DOLoRes               & 48       & 48        & 74        &  74       & 69        & 64        & --        & --        & --        & --        & --        & --        & --        & --        & --        & --   \\
        Wendelstein 2-m/WWFI      & --       & --        & --        & --        & --        & --        & --        & --        & 45        & 44        & 1619      & 1245      & 41        & 38        & --        & --   \\
        OSN 1.5-m/CCD             & --       & --        & --        & --        & 1499      & 1473      & --        & --        & --        & --        & --        & --        & --        & --        & --        & --   \\
        INT/WFC                   & 37       & 34        & --        & --        & --        & --        & 2         & 2         & --        & --        & 90        & 86        & --        & --        & --        & --   \\
        BTA/SCORPIO2              & --       & --        & --        & --        & --        & --        & --        & --        & 6         & 0         & 15        & 13        & --        & --        & --        & --   \\
        LCOGT/SBIG                & --       & --        & --        & --        & --        & --        & --        & --        & --        & --        & 51        & 42        & --        & --        & --        & --   \\
        OSN 0.9-m/CCD             & --       & --        & --        & --        & 78        & 78        & --        & --        & --        & --        & --        & --        & --        & --        & --        & --   \\
        LCOGT/Sinistro            & --       & --        & --        & --        & --        & --        & --        & --        & --        & --        & 42        & 36        & --        & --        & --        & --   \\
        Gemini N/GMOS             & --       & --        & --        & --        & --        & --        & --        & --        & 17        & 17        & 42        & 42        & 12        & 12        & 12        & 12   \\
        IRTF/MORIS                & --       & --        & --        & --        & --        & --        & --        & --        & --        & --        & 113       & 0         & --        & --        & --        & --   \\
        \hline
    \end{tabular}
\end{table*}

\begin{figure*}
	\includegraphics[width=\textwidth]{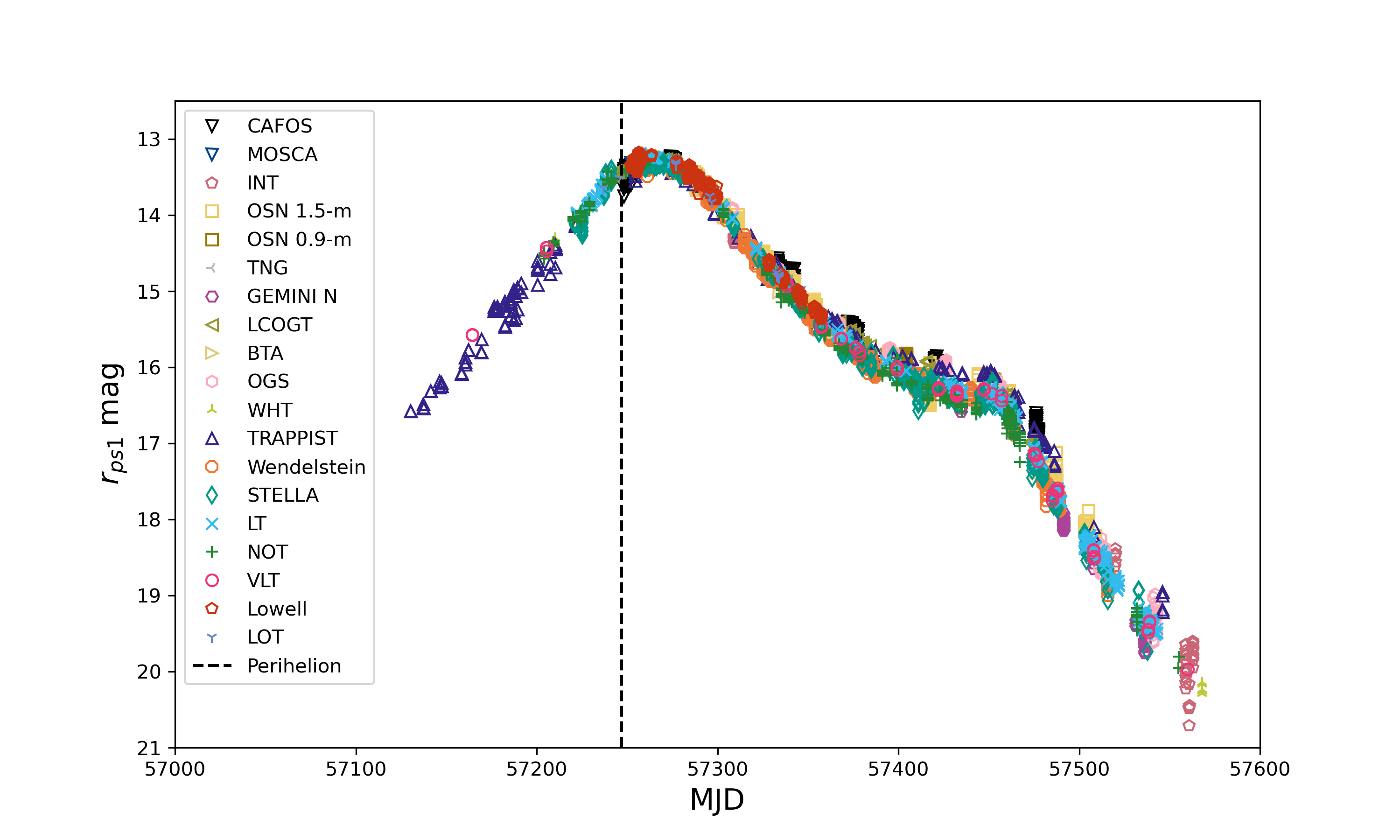}
    \caption{Light curve of 67P/Churyumov-Gerasimenko measured within 10,000 km aperture. Photometry has been calibrated and scaled to the PS1 \textit{r}-band. The vertical dotted line shows time of perihelion on 2015 August 13. \amend{The photometry behind this plot can be found in Table~\ref{tab:photometry}.}}
    \label{fig:lightcurve}
\end{figure*}

\subsection{Summary}
Figure~\ref{fig:lightcurve} shows the \textit{r}-band light curve of 67P, calibrated to the PS1 photometric system, compiling all the data processed through our pipeline. Table~\ref{tab:Results} present a summary of the data processed. In this sub-section we briefly describe specific results from specific data sets.

The VLT provided a high quality data set that ran with few issues. As such it was used as the test data for the initial development of the pipeline. The VLT data also helped us constrain the limitations of the automatic detection, with its wide observing window it observed the comet at its faintest. From this we determined that the pipeline works best when observing a comet brighter than 20 mag.

The regular observations by the LT were well suited for our pipeline producing well calibrated and consistent results in \textit{g}-, \textit{r}-, \textit{i}- and \textit{z}-bands. These data formed the backbone of our comparison and we used this as a "true" representation of the light curve. The colours obtained from the LT were also used as our starting point when approaching colour calibration of the remaining data.

The NOT/ALFOSC data presented an issue; noisy edges left over from the data reduction process. These noisy edges were sometimes incorrectly identified as sources by Astrometry.net causing it to fail to solve. We initially tried masking the edges but the problem persisted even after masking. We concluded that then it must have been an issue with the fields of stars themselves. We took the decision to discard these images rather than adapt the pipeline to mask the specific noise pattern since this affected only around 12 per cent of the images in the NOT data, and even fewer in the data set as a whole. Despite this, it is one of the most well calibrated and comprehensive data sets in both \textit{r}- and \textit{g}-bands. 32 images taken by CAFOS on the Calar Alto Observatory (CA) 2.2-m telescope had noisy artefacts, similar to the NOT, which caused the pipeline to fail. Again, we took the decision to simply discard these frames. The calibration of some of these data has wide variations within nights due to the small number of calibration stars within the field of view. This led to some differing estimations of the zero point in each frame as \textsc{calviacat} tried to fit a line to a small number of points. The large number of exposures taken each night allowed us to remove outliers in the calibration. While this data set is large it is concentrated on small groups of consecutive nights separated by weeks rather than long-term monitoring.

The Lowell data set has good coverage around perihelion, but somewhat inconsistent calibration due to varying quality between frames. Several frames contained dead pixels which sometimes would lie on top of a star, making the calibration less accurate. The 43 frames where the dead pixels lay within the comet aperture were discounted. 

The European Space Agency (ESA) Optical Ground Station (OGS) data contained 18 frames which were discounted due to being pointed towards the wrong area of the sky. Another 37 frames did not have enough background stars to perform calibration.

The pipeline failed to run on Infrared Telescope Facility (IRTF), New Technology Telescope (NTT) and NOT/STANCAM data due to the small field of view in the images. There were few stars within the field which meant astrometry and photometry calibrations failed.

It is worth mentioning the William Herschel Telescope (WHT), LULIN Observatory One-Meter Telescope (LOT), Telescopio Nazionale Galileo (TNG), Isaac Newton Telescope (INT) and Bolshoi Teleskop Alt-azimutalnyi (BTA). They are well calibrated data but have sparse coverage having only 3 to 5 nights of observations in each set. The consistent calibrations made possible due to the larger data set they are a part of meant that they are still useful in the final data to fill in gaps and aid in confirming outbursts.

Any data not mentioned above ran through our pipeline successfully and was generally well calibrated. A common issue in these data were the occasional lack of background stars to use for photometric calibration but this typically affected less than 10 per cent of images.

Overall the pipeline worked well and processed the majority ($\sim$ 83 per cent) of the data and produced well calibrated and consistent results across the different data sets. The pipeline works best when the comet is brighter than mag 20 brightness and in a well-exposed, but not too crowded, field of stars. Without these conditions the comet identification and calibration becomes increasingly inaccurate.

A limitation highlighted by the NOT/ALFOSC and CA/CAFOS data is that the pipeline has no way of adjusting for any noisy edges or artefacts that may remain after data reduction. This noise often was misidentified as sources by \textsc{SExtractor} and Astrometry.net which caused them either to fail or give inaccurate results. The other limitation is the pipeline only does simple aperture photometry around the comet and does not take into account any contribution of the background flux from nearby bright stars that are outwith the aperture. We decided against implementing a fix for both of these issues because of the small proportion of images they affect.

\subsection{Offset between telescopes around low phase angles}
\label{sec:offsets}
\begin{figure}
	\includegraphics[width=\columnwidth]{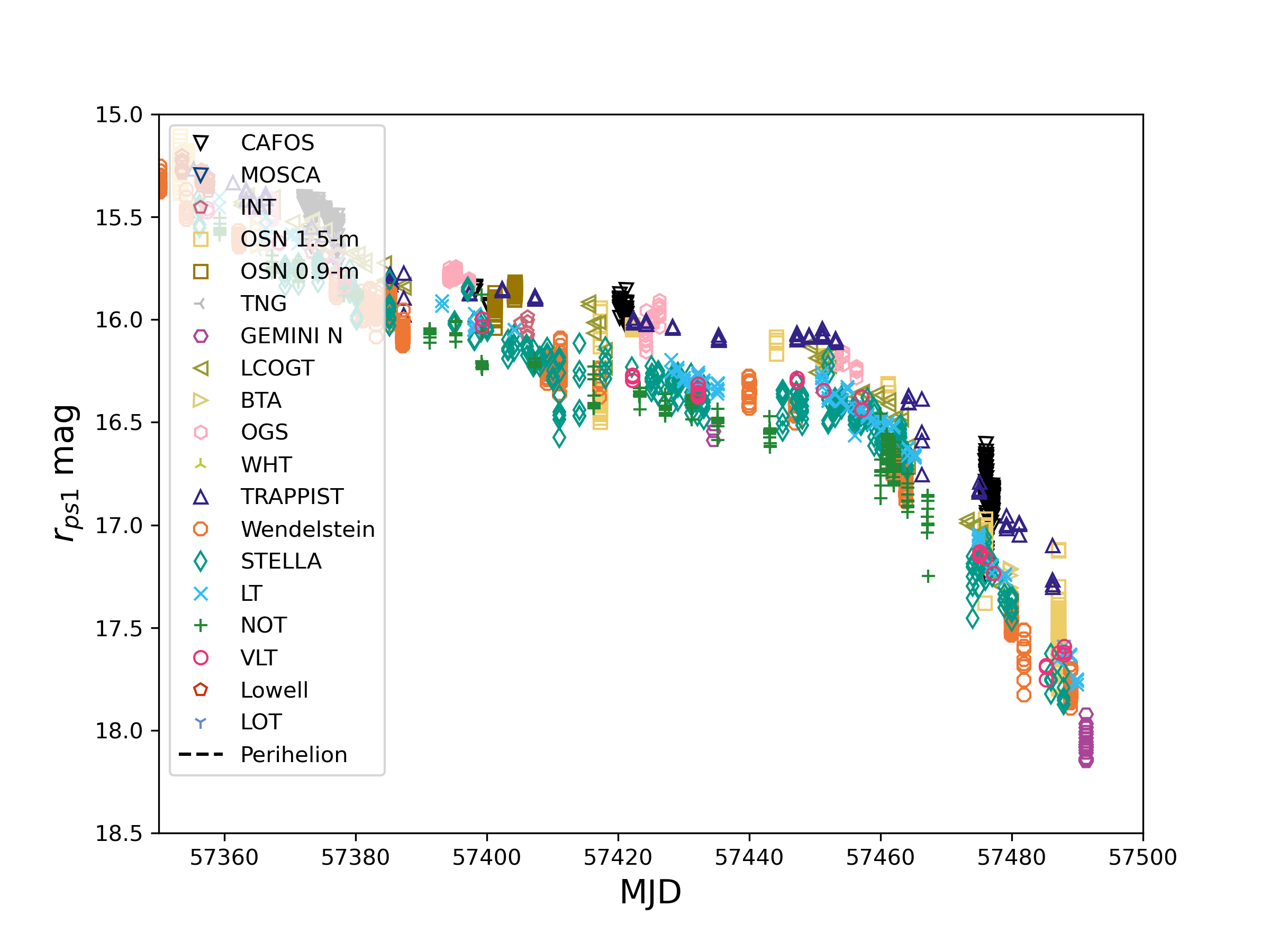}
    \caption{Light curve around low phase angles. All points have been calibrated to \textit{r} PS1 band. A clear offset is seen in the TRAPPIST (blue triangles), OSN (yellow squares), OGS (pink hexagons) and CAFOS (black downward-triangles) points which were measured in Johnson/Cousin \textit{R} filter. The majority of the other points were measured in sloan-\textit{r} type filters.}
    \label{fig:offset}
\end{figure}

A peculiar effect we see in our data is a significant shift in \textit{r}-band magnitudes at low phase angles between different telescopes (Figure~\ref{fig:offset}). The TRAPPIST data is the best example of this, it follows the overall brightness trend but is shifted relative to the curve around low phase angles, the bump in the light curve around MJD 57450 when the comet was at opposition in 2016.  Initially we thought it was an effect of the slight bandpass differences between the Sloan-\textit{r} and Johnson/Cousins \textit{R} filters. Some other data sets that use Johnson/Cousins filters, e.g. OSN, OGS and CAFOS, appear to align better with TRAPPIST however this is not true of all data taken in this filter, for example the NOT data does not have an offset and follows the trend of the majority of the data. This offset persisted after colour calibration. We looked at \textit{r--i} colours around low phase angles to see if there was a change in the colour but this was not seen, see section~\ref{sec:colour}. We also investigated if there was any correlation between colour and geocentric distance, airmass or seeing. We did not see any correlation so it remains a mystery as to what is causing this offset. In order to aid in meaningful comparisons between the data we needed to correct these offsets and line the points up with the rest of the curve. To correct for these offsets we first subtracted the overall trend of the light curve leaving us with a scatter of points around the average. For each data set that was offset from the average, we modelled the offset as a function of time using a simple straight-line fit. Each fit was then subtracted from their respective data so the averages of each data set followed the average of the overall curve.

\section{Analysis}
\subsection{Coma Colour}
\label{sec:colour}
\begin{table}
    \centering
    \caption{Average colour of 67P coma across the observation period measured by different instruments. All instruments are calibrated to the Pan-STARRS magnitude system.}
    \label{tab:colours}
    \begin{tabular}{lll}
        \hline
        Telescope/Instrument               & filter range & colour index  \\
        \hline
        Gemini N/GMOS                      & \textit{g--r}                  & 0.65 $\pm$ 0.04 \\
                                           & \textit{r--i}                  & 0.26 $\pm$ 0.03  \\
                                           & \textit{i--z}                  & 0.03 $\pm$ 0.06  \\
        LT/IO:O                            & \textit{g--r}                  & 0.61 $\pm$ 0.004\\
                                           & \textit{r--i}                  & 0.27 $\pm$ 0.004 \\
                                           & \textit{i--z}                  & 0.08 $\pm$ 0.01  \\
        NOT/ALFOSC                         & \textit{g--r} (from \textit{V} and \textit{R})   & 0.61 $\pm$ 0.004\\
        STELLA/WIFSIP                      & \textit{g--r}                  & 0.64 $\pm$ 0.02 \\
                                           & \textit{r--i}                  & 0.24 $\pm$ 0.03  \\
                                           & \textit{i--z}                  & 0.08 $\pm$ 0.05  \\
        TRAPPIST-South/CCD                 & \textit{g--r} (from \textit{B},\textit{V} and \textit{R}) & 0.60 $\pm$ 0.004 \\
                                           & \textit{r--i} (from \textit{R} and \textit{I})   & 0.20 $\pm$ 0.004 \\  
        Wendelstein 2-m/WWFI               & \textit{g--r}                  & 0.59 $\pm$ 0.02  \\
                                           & \textit{r--i}                  & 0.22 $\pm$ 0.02  \\    
        \hline
    \end{tabular}
\end{table}

\begin{figure*}
	\includegraphics[width=\textwidth]{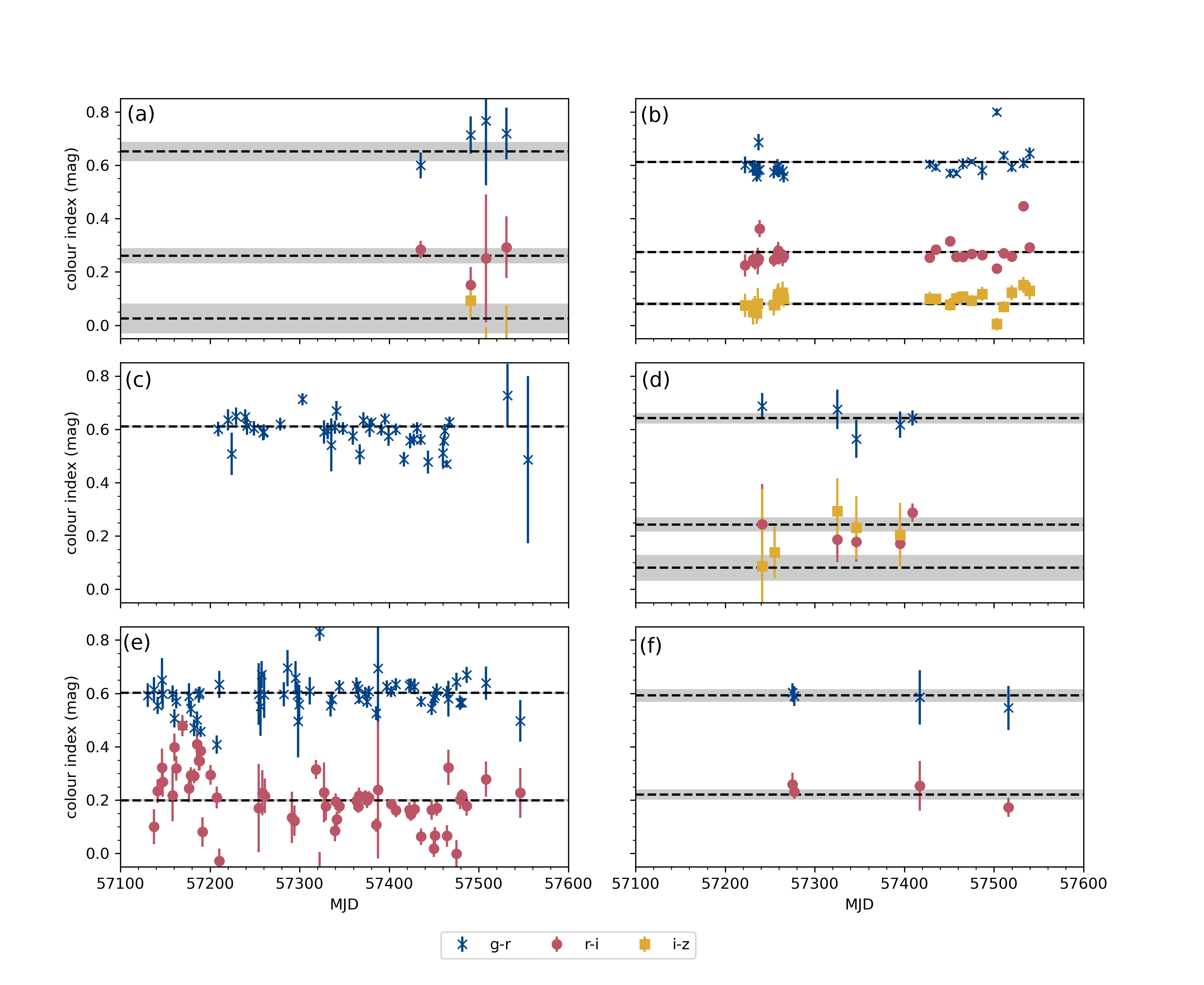}
    \caption{Colour against modified Julian date between 2015 March 19 and 2016 July 31 for comet 67P/Churyumov-Gerasimenko measured with (a) Gemini-North, (b) the Liverpool Telescope, (c) the Nordic Optical Telescope, (d) STELLA, (e) TRAPPIST-South and (f) the Wendelstein 2-m telescope in the \textit{g-, r-, i-} and \textit{z}-bands. The weighted mean colours are shown with the horizontal dotted lines and uncertainties as the shaded areas.}
    \label{fig:colour}
\end{figure*}

The coma colour remains more or less constant throughout the apparition (Figure~\ref{fig:colour}) indicating no significant change in the gas production relative to dust production around perihelion, which would be expected to cause a decrease in \textit{g--r}, for example. Table ~\ref{tab:colours} summarises the average colours measured by six different instruments during the campaign in \textit{g--r}, \textit{r--i} and \textit{i--z}. The colours for the NOT/ALFOSC and TRAPPIST-South have been converted from \textit{B, V, R, I} to \textit{g, r, i} \citep{Jester2005}. The \textit{g--r} colour of 67P is consistent with what we would expect the dust from a comet to look like at these heliocentric distances \citep{Jewitt1986}. \citet{Boehnhardt2016} reports a minor \textit{g--r} colour change from 0.56 to 0.62 measured within a 10,000 km apertures between 2015 September 10 and 2016 May 7. They report this from a sample of just four data points from across their data. We do not see the same trend in our calibrations of the Wendelstein data, in fact we see an opposite trend, however this discrepancy could be explained by the differences in calibration methods between our works. Since we have access to colour data from a much wider span of time we can say that we do not see this subtle colour change in any of our data. 

\begin{figure}
	\includegraphics[width=\columnwidth]{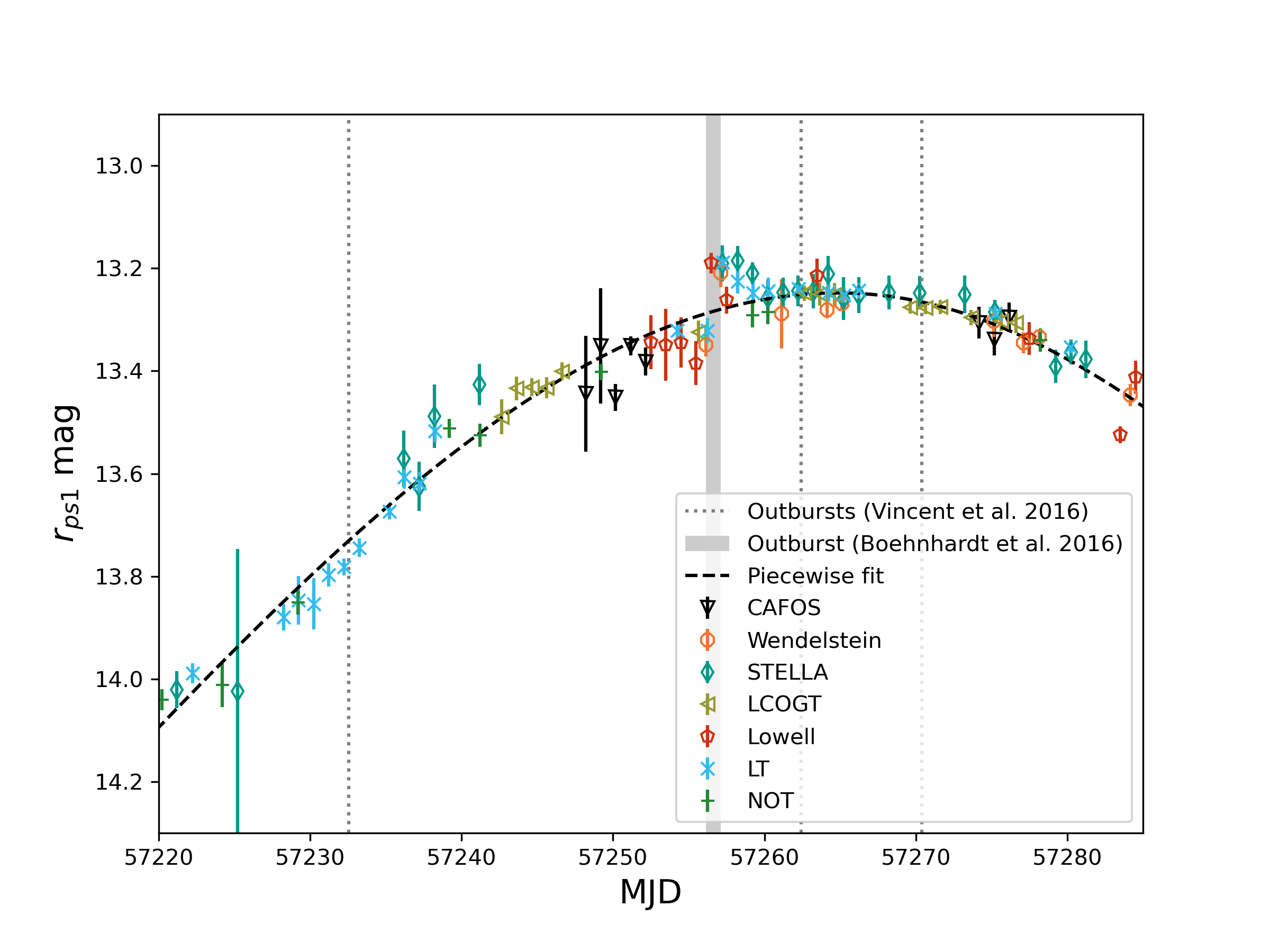}
    \caption{Light curve of 67P/Churyumov-Gerasimenko around perihelion between 2015 July 17 and 2015 September 25. Photometry has been calibrated and scaled to the \textit{r}-band. A piecewise fit trend line has been plotted and the time of the outburst seen in \citet{Boehnhardt2016} has been highlighted. The grey dotted line show the times of the brightest outbursts seen by \citet{Vincent2016}.}
    \label{fig:perihelion}
\end{figure}

\begin{figure*}
	\includegraphics[width=\textwidth]{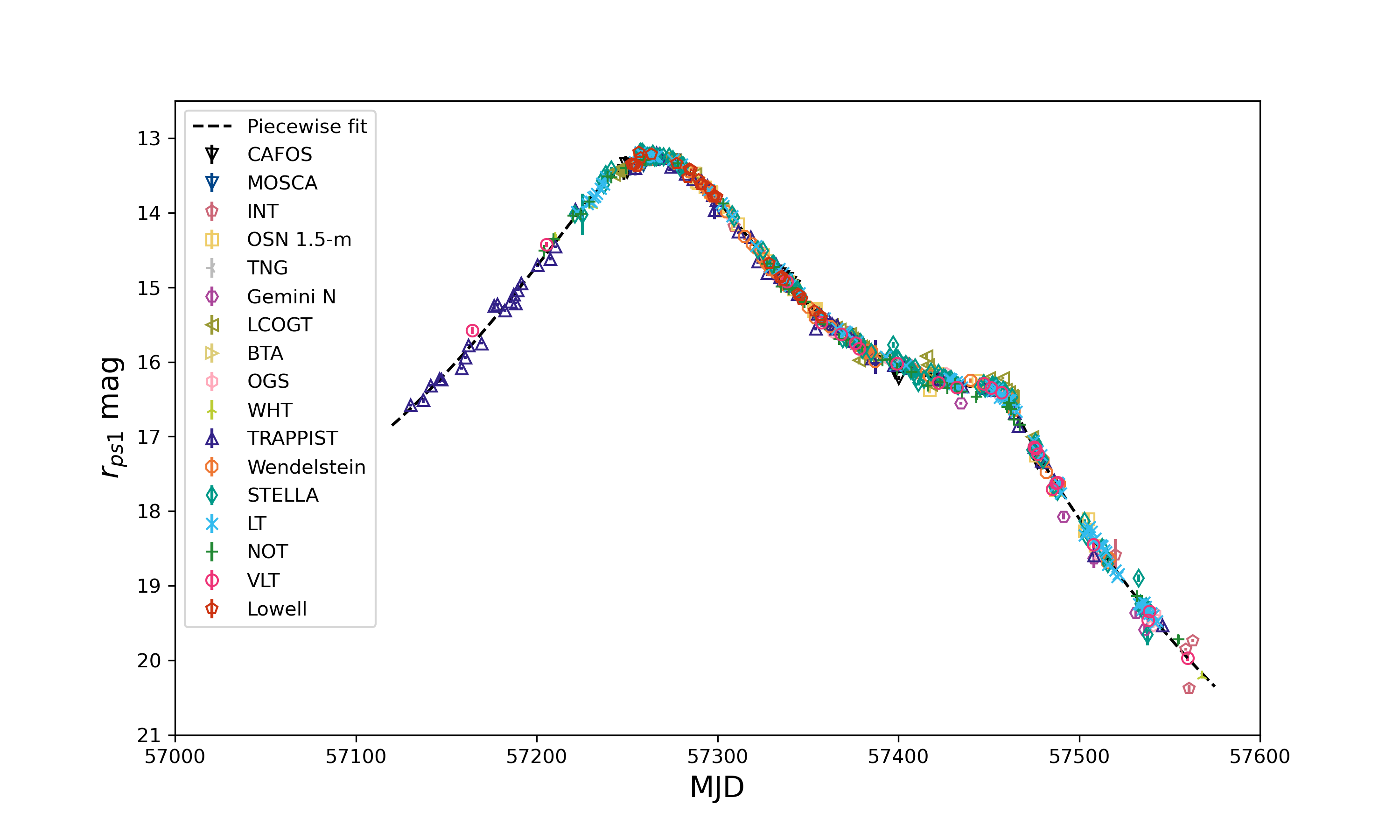}
    \caption{Light curve of 67P/Churyumov-Gerasimenko measured within 10,000 km aperture. Photometry has been calibrated and scaled to the \textit{r}-band. A piecewise fit trend line has been plotted. \amend{The photometry behind this plot can be found in Table~\ref{tab:nightly_average}.}}
    \label{fig:lightcurve_fit}
\end{figure*}

\subsection{Searching for Outbursts}
67P was observed in \textit{r}-band filters for almost its entire perihelion passage. We measured the maximum brightness of the comet as $\sim$13.2 within a 10,000 km aperture for the period of 2015 late August to early September. The light curve (Figure~\ref{fig:lightcurve}) follows the predictions \citep{Snodgrass2013} well and does not show any large-scale deviations from the expectations, which indicates the activity level remained more or less constant between apparitions. A brightness increase of $\sim$0.14 mag was obvious in multiple data sets on 2015 August 22, indicated in Figure~\ref{fig:perihelion}, confirming the outburst seen by \citet{Boehnhardt2016} with the Wendelstein telescope. The number of telescopes pointed at 67P that night allowed us to constrain the event to within a few hours. The last observation taken by the LT at 05:51:25 UTC measured a brightness of 13.34$\pm$0.03, then about five hours later it was observed by the Lowell telescope between 11:17:19 and 11:46:24 UTC which measured an average brightness of 13.20$\pm$0.02. This increase in brightness is seen by the LT and Wendelstein the following night. LT measured 13.19$\pm$0.02 and Wendelstein measured 13.22$\pm$0.03.

\begin{figure}
	\includegraphics[width=\columnwidth]{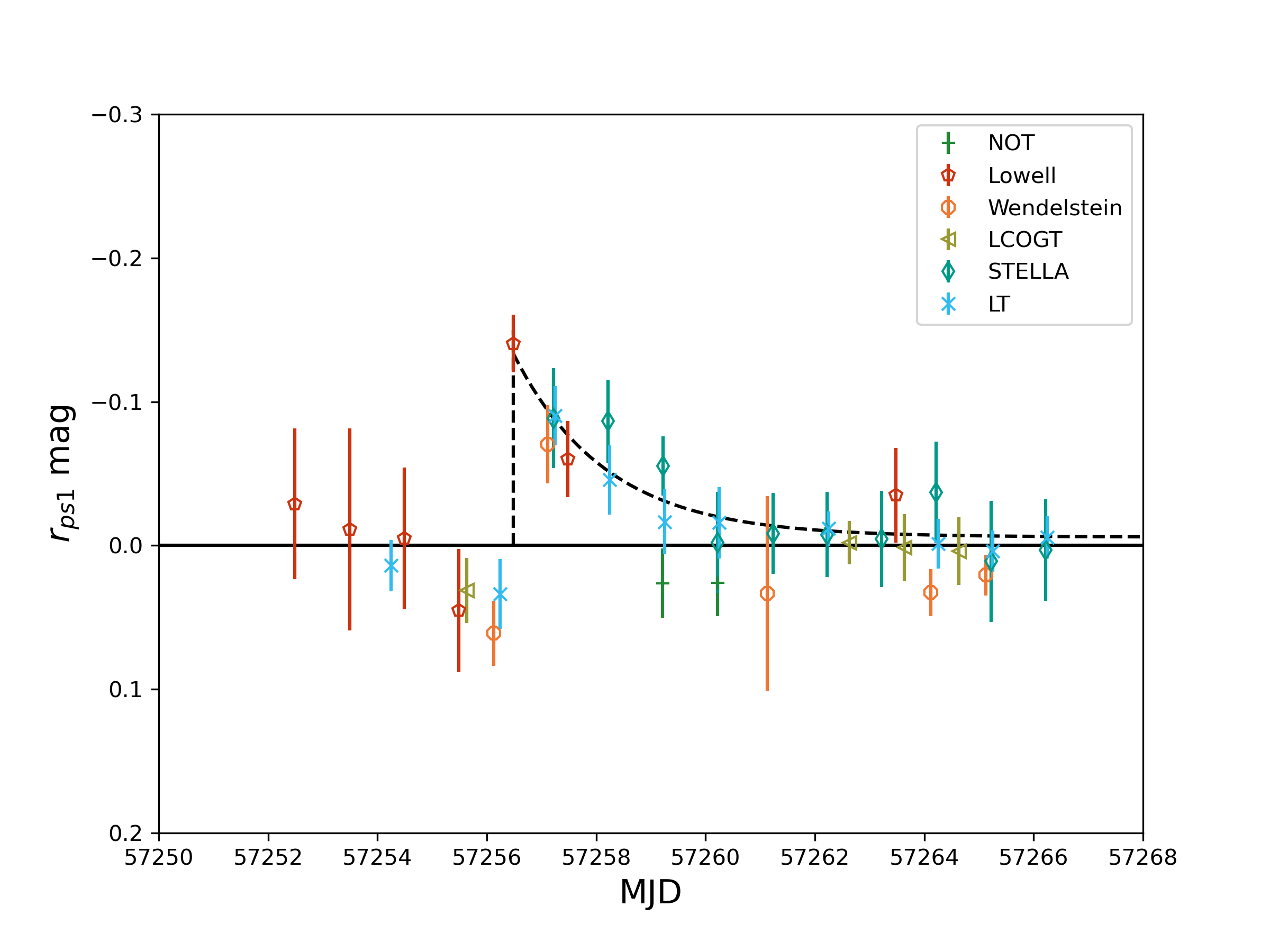}
    \caption{Light curve around 2015 August 22 with the baseline photometric trend removed. An anomalous increase in the brightness is obvious. The Lowell points have been shifted to match the trend. The anomaly shows signs of being an outburst with a rapid brightening with an exponential fall-off. We estimate a brightening of 0.14 \amend{mag}.}
    \label{fig:outburst}
\end{figure}

In order to properly characterise these outbursts and discover others missed by manual inspection we removed the underlying photometric trend. We modelled the trend as a simple polynomial piecewise fit. The data are scaled and shifted to fit to the curve as described in section~\ref{sec:offsets}. This was done because of subtle offsets between the data sets remaining after the colour calibration. Figure~\ref{fig:lightcurve_fit} shows the light curve with the offsets between data removed and the piecewise fit plotted underneath. Figure~\ref{fig:outburst} shows the outburst of 2015 August 22 with the trend removed. We modelled an exponential fit to the outburst, peaking at 0.14 $\pm$ 0.02 mag brighter than the baseline and falling off as $m\propto~e^{-0.59t}$ \amend{, where t is measured in days}. No further outbursts were seen after the removal of the baseline trend. We tried to find evidence of outbursts \amend{that were seen in situ} on 2016 February 19 \citep{Grun2016} and 2016 July 3 \citep{Agarwal2017} but we couldn't find anything convincing. \amend{We also looked to see if we could confirm the potential outburst seen from the ground on 2015 September 19 \citep{Knight2017} but we could not find convincing evidence of brightening within the other data sets at this time.}

\section{Discussion}
\subsection{Outburst of 2015 August 22}
We can confirm the outburst seen in the comet coma morphology by \citet{Boehnhardt2016} in our analysis of the 67P photometry. The date and time of this event corresponds to an outburst seen by the NavCam instrument onboard Rosetta: outburst \#16 from \citet{Vincent2016}, which was observed on 2015 August 22 06:47:04 UTC. Outburst \#16 could be connected to our outburst, it is bright and occurs immediately before the brightness increase seen in the ground-based data at 2015 August 22 11:17:19 UTC. The data taken with NavCam is uncalibrated so we do not know exactly how bright it is compared to the other outbursts seen with OSIRIS. Judging the images by eye we can see that the outburst is as bright as, if not brighter than, the other outbursts and shares morphological similarities with the brightest outbursts seen with OSIRIS. 

Looking at NavCam images from ESA's Planetary Science Archive (PSA), the first image taken $\sim$2 hours following the outburst seems to show increased activity; it has an increased brightness in the inner coma compared with other images taken around that time. This indicates a possible longer term event compared to typical events seen from the spacecraft, which appear only in single frames. However, the comet was observed from a different orientation in the image following the outburst, and activity level varies depending on the part of the surface that is illuminated so it is difficult to make a direct comparison and make a clear statement about the longevity of the outburst. OSIRIS did not acquire images at the time of the outburst, the images that were closest in time to the outburst were taken 2016 August 22 05:55:43 UTC and 2016 August 23 08:20:06 UTC. These images taken before and after the outburst do not show any significantly increased activity (C Tubiana 2022, personal communication, 22 July).

The Boehnhardt outburst looks different in morphology to the Vincent outburst, the former is a jet-like structure while the latter is much broader and fan-like in its appearance, although it is essential to point out that these two structures are very different in scale. The outbursts photographed by Rosetta are of the order of 10 km in size whereas the Boehnhart event is approximately 5,000 km in length. The source locations estimated for these events also differ, \citet{Boehnhardt2016} suggests the feature originated from latitudes between +5$^{\circ}$ and +10$^{\circ}$ on the nucleus whereas \citet{Vincent2016} see their outburst coming from a latitude of -40$^{\circ}$. \amend{This discrepancy may be due to differing coordinate systems, since \citet{Boehnhardt2016} uses a simplified spherical model to estimate the source location and \citet{Vincent2016} uses more accurate planetographic coordinates. However, we believe the differences between these systems are not enough to explain the large offset in source latitudes.} Since the scales of these outbursts differ by orders of magnitude, it could be possible that outburst \#16 is but one of many small outbursts that contribute to this larger coma change. Outburst \#15 \citep{Vincent2016} was seen about 24 hrs before \#16 and could be contributing to the brightening but there was no brightening seen in the data when outburst \#15 happened. It could be possible a larger outbursting event was missed by the in situ instruments, however this is unlikely since during that time the probe was regularly monitoring the nucleus, taking images with an average separation of 12 minutes and some as short as 5 minutes \citep{Vincent2016}. All of this uncertainty makes it difficult to draw a definitive connection between the Boehnhardt event and in situ observations.

\subsection{Searching for other confirmed outbursts and linking observations to surface changes}
The outburst of 2015 August 22 is on the smaller side of outbursts typically seen from the ground in other comets. While this outburst was easily spotted, it's very possible it could have been missed had we not known where to look. It was noticed due to its connection with the morphology change seen by \citet{Boehnhardt2016}. Other outbursts seen by the spacecraft were not seen on the ground. \citet{Grun2016} reports a sustained increase in Af$\rho$, using a 5000 km aperture, of the comet around the event of 2016 February 19 (MJD 57437.4), based on TRAPPIST data. However this sustained brightness increase is not seen in our magnitude data.  This could be due to the low phase angle at the time of observations, meaning the phase angle effects masked any potential signal from the data. It is worth noting that our magnitudes are not phase-corrected whereas \citet{Grun2016} presents phase corrected data. Also this sustained brightness could have been subtracted from the data during detrending. We do not detect the brightening independent of the spacecraft data.

A major goal of this study was to see if any surface changes seen by Rosetta could be connected to observations made from the ground. One of the most notable surface changes on the comet was the Aswan cliff collapse \citep{Pajola2017}. This collapse was linked to a bright outburst seen by NavCam on 2015 July 10 (MJD 57213.1). There is unfortunately a gap of several days in the ground-based data that coincides with this event meaning any increase in brightness would have been missed so it is impossible to say if this event could have been visible from the ground. \citet{El-Maarry2019} summarises and maps the major nuclear surface changes observed by Rosetta. We compared these observations to the estimated source positions of both the Vincent and Boehnhart outbursts but we find no obvious signs of a surface change that corresponded to either position. 

\subsection{Dust mass estimate}
In order to create more meaningful comparisons to physical quantities we estimate the mass of the 2015 August 22 outburst. We assume a dust \textit{r}-band geometric albedo of 4.17 per cent and a total geometric cross-sectional area, $G$, defined in \citet{Kelley2021a} as:

\begin{equation}
    G=\frac{\pi r_{\mathrm{h}}^2\Delta^2}{A_p\Phi(\theta)}10^{-0.4(m-m_{\odot})}
\end{equation}

where $\Delta$ is the observer-comet distance, $\Phi(\theta)$ is the coma phase function evaluated at phase angle $\theta$ \citep{Schleicher2010}, $m$ is the apparent magnitude of the total dust coma and $m_\odot$ is the apparent magnitude of the Sun at 1 au in the same bandpass and magnitude system. In order to convert $G$ into dust mass we need to make an assumption of the grain density and grain size distribution. We assume a grain density of 500 kg m$^{-3}$ \citep{Jorda2016} and a grain size distribution of $dn/da = a^{-2.6}$ \citep{Vincent2016}, we constrain the dust grain radii between \amend{1 $\mu$m and 10 $\mu$m}. To calculate the mass of the outburst we subtract the coma mass after the outburst from the coma mass before the outburst. Using these we estimate the mass of our outburst to be \amend{2.0$\times 10^5$ kg} (\amend{$\sim$12} per cent of the total coma), which puts it in agreement with the mass estimates made by \citet{Vincent2016}, who puts a constraint of $10^4$ kg on the typical dust mass of outbursts seen by Rosetta, with $10^5$ kg being the largest seen. \citet{Grun2016} claim to observe an outburst of mass $10^3$ kg from the ground, with such sensitivity we would expect to see many more outbursts than we do. Our estimate is also in line with similar scale outbursts seen on other comets observed from the ground; \citet{Kelley2021a} estimates that the mass of outbursts from 46P/Wirtanen lie between 3$\times 10^4$ kg to \amend{5 $\times 10^6$} kg. It is encouraging to see that the outburst we see is of the same mass as the ones seen by Rosetta as this implies that the outburst seen from the ground is, if not one observed directly by Rosetta, an outburst of a similar scale to the largest seen. However, this raises the question as to why apparently none of the other similarly large outbursts were seen from the ground. The other brightest outbursts, including outburst \#12, the brightest seen by \citet{Vincent2016}, go unseen in the photometry.  Perhaps this is due to the challenging viewing conditions that were present during the early part of the campaign when 67P was most active. 

\subsection{Comparison to 2021--2022 apparition}
\label{sec:2021}
We made a comparison to data taken taken during the 2021--2022 apparition, where the viewing conditions were a lot more favourable. Despite the better viewing conditions, there wasn't as an intensive monitoring campaign for this apparition. \citet{Sharma2021} observed two outbursts using the 70-cm GROWTH-India Telescope on 2021 October 29 and 2021 November 17. The second outburst was also observed and confirmed by the LCO Outbursting Objects Key Project \citep{Kelley2021b, Lister2022}. These outbursts came 4 days before and 15 days after perihelion respectively. This lends credence to the fact that activity levels remain similar between orbits since it was at this point around perihelion in the 2015--2016 apparition where Rosetta detected the highest rate of outbursts. The outburst we saw on 2015 August 22 occurred 9 days after perihelion. The Sharma outbursts were measured to have masses of 5.3 $\times 10^5$ kg and \amend{1.3 $\times 10^6$} kg respectively. This is consistent with the mass we estimated for our outburst but it is still an order of magnitude larger than the typical outburst seen by Rosetta.  

\begin{figure}
	\includegraphics[width=\columnwidth]{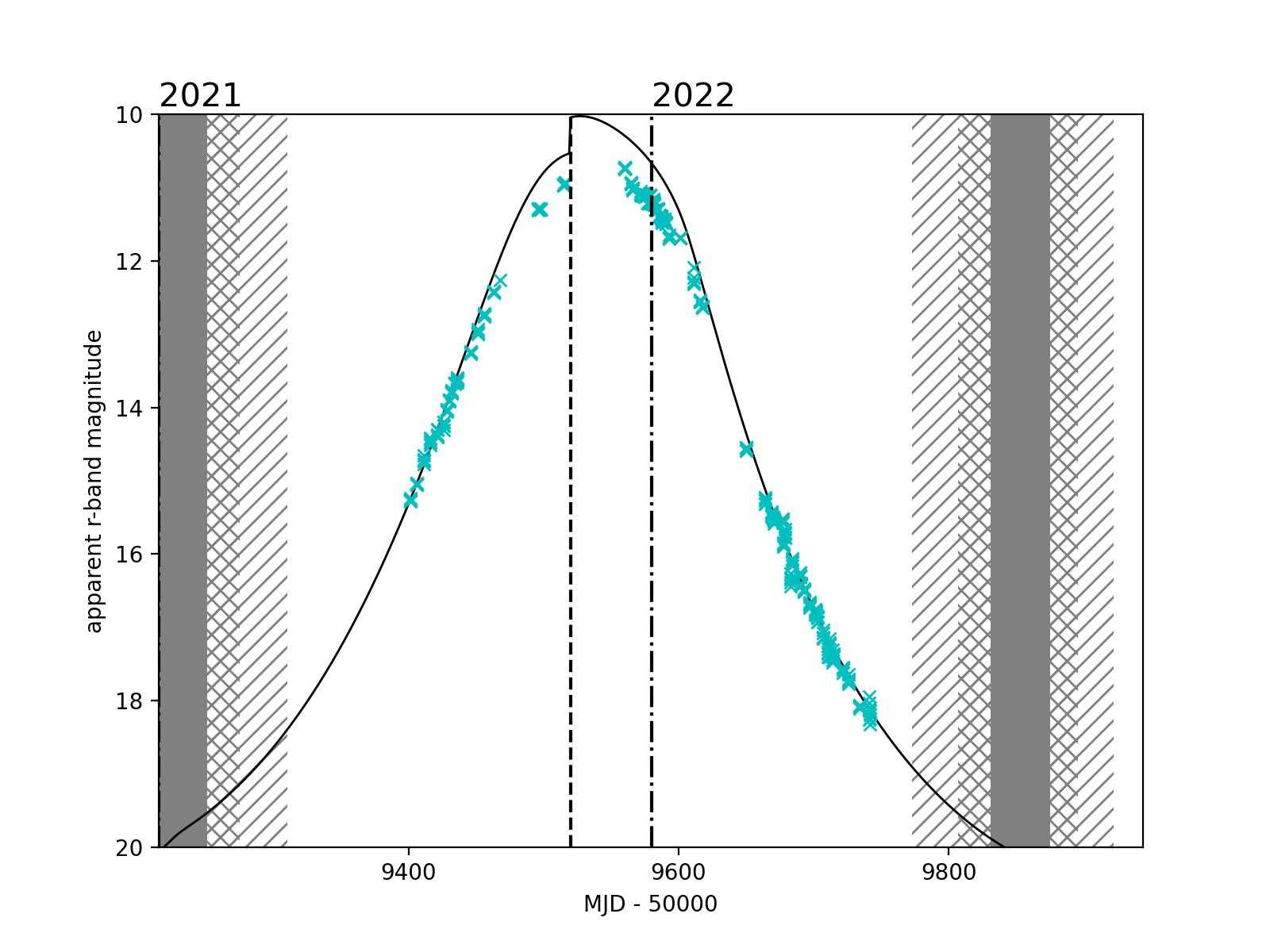}
    \caption{\textit{r}-band light curve of the 2021--2022 67P apparition. Observations were taken at the LT. The solid black curve is the predicted brightness of the comet. The step in the curve between pre- and post-perihelion is not a real feature and is due to the simplistic power law functions used in the prediction. The vertical dashed line shows the time of perihelion. The hatched patterns shows when solar elongation below 50$^{\circ}$ (hatched), 30$^{\circ}$ (cross-hatched) and 15$^{\circ}$ (solid grey). The dash-dotted line shows the division between years. \amend{The photometry behind this plot can be found in Table~\ref{tab:LT2022}.}}
    \label{fig:LT2021}
\end{figure}

We observed 67P with the LT during its 2021--2022 apparition, the \textit{r}-band light curve is shown in Figure~\ref{fig:LT2021}. As with the previous apparition, the comet follows the predictions well. The predictions were created using the same method as \citet{Snodgrass2013}. The discrepancy around perihelion is due to the simplification of the models and should not be seen as a deviation from predictions. The data is a very good match for the predictions pre- and post-perihelion which suggests no difference in activity levels. No outbursts were seen in this data. This data was unfortunately marred by extended periods of telescope downtime due to a volcanic eruption on La Palma. As luck would have it, the eruption coincided with perihelion and the two outbursts that were seen by \cite{Sharma2021}, which meant we were unable to independently confirm these outbursts using the LT. 

All of this highlights the fact that characterising small-scale outbursts of a comet and linking it to nuclear activity is still challenging, even when we have direct comparisons from spacecraft data.

\section{Conclusions}

We developed a pipeline for the consistent calibration of the multitude of disparate data from the ground-based observing campaign accompanying the Rosetta mission. The pipeline worked well with a processing success rate of $\sim$83 per cent across the data. The calibrated data allowed for a careful search for outbursts through the perihelion period between 2015 April and 2016 August. We discovered one outburst on 2015 August 22 with a magnitude increase of $\sim$0.14 mag. This event confirms that the brightening seen in \citet{Boehnhardt2016} was a sign of an outburst. Linking this event with in situ outbursts proved challenging: while an in situ outburst was seen within the same time period as the brightening event, discrepancies in the surface origin estimates and the differences in scale of the in situ outbursts compared to the large-scale coma morphology made it hard to prove that there was a direct link between them. No other outbursts were seen in our data despite the many in situ events observed. We conclude that events of this scale are extremely challenging to observe from the ground and bridging the gap between large-scale coma changes and small-scale nuclear activity remains to be understood. 

\section*{Acknowledgements}

This work was supported by the UK Science and Technology Facilities Council. We would like to thank all the observers who were involved in the original observation campaign. They are too numerous to list here but you can find the complete list in \citet{Snodgrass2017}. Thanks to Cecilia Tubiana for her assistance in the acquisition and comparison of archival OSIRIS and NavCam data. \amend{We thank the reviewer for their helpful and constructive feedback. For the purpose of open access, the author has applied a Creative Commons Attribution (CC BY) licence to any Author Accepted Manuscript version arising from this submission.}

\section*{Data Availability}
 
All of the original imaging data from the ground-based observing campaign will be made publicly available at the ESA Planetary Science Archive (http://www.cosmos.esa.int/web/psa). In addition, much of the raw data are available from individual observatory archive facilities.



\bibliographystyle{mnras}
\bibliography{references} 

\begin{thebibliography}{}
\makeatletter
\relax
\def\mn@urlcharsother{\let\do\@makeother \do\$\do\&\do\#\do\^\do\_\do\%\do\~}
\def\mn@doi{\begingroup\mn@urlcharsother \@ifnextchar [ {\mn@doi@}
  {\mn@doi@[]}}
\def\mn@doi@[#1]#2{\def\@tempa{#1}\ifx\@tempa\@empty \href
  {http://dx.doi.org/#2} {doi:#2}\else \href {http://dx.doi.org/#2} {#1}\fi
  \endgroup}
\def\mn@eprint#1#2{\mn@eprint@#1:#2::\@nil}
\def\mn@eprint@arXiv#1{\href {http://arxiv.org/abs/#1} {{\tt arXiv:#1}}}
\def\mn@eprint@dblp#1{\href {http://dblp.uni-trier.de/rec/bibtex/#1.xml}
  {dblp:#1}}
\def\mn@eprint@#1:#2:#3:#4\@nil{\def\@tempa {#1}\def\@tempb {#2}\def\@tempc
  {#3}\ifx \@tempc \@empty \let \@tempc \@tempb \let \@tempb \@tempa \fi \ifx
  \@tempb \@empty \def\@tempb {arXiv}\fi \@ifundefined
  {mn@eprint@\@tempb}{\@tempb:\@tempc}{\expandafter \expandafter \csname
  mn@eprint@\@tempb\endcsname \expandafter{\@tempc}}}

\bibitem[\protect\citeauthoryear{Agarwal et~al.,}{Agarwal
  et~al.}{2017}]{Agarwal2017}
Agarwal J.,  et~al., 2017, \mn@doi [\mnras] {10.1093/mnras/stx2386}, 469, s606

\bibitem[\protect\citeauthoryear{{Bertin} \& {Arnouts}}{{Bertin} \&
  {Arnouts}}{1996}]{Bertin1996}
{Bertin} E.,  {Arnouts} S.,  1996, \mn@doi [A\&AS] {10.1051/aas:1996164}, \href
  {https://ui.adsabs.harvard.edu/abs/1996A&AS..117..393B} {117, 393}

\bibitem[\protect\citeauthoryear{Boehnhardt, Riffeser, Kluge, Ries, Schmidt  \&
  Hopp}{Boehnhardt et~al.}{2016}]{Boehnhardt2016}
Boehnhardt H.,  Riffeser A.,  Kluge M.,  Ries C.,  Schmidt M.,   Hopp U.,
  2016, \mn@doi [\mnras] {10.1093/mnras/stw2859}, 462, S376

\bibitem[\protect\citeauthoryear{Choukroun et~al.,}{Choukroun
  et~al.}{2020}]{Choukroun2020}
Choukroun M.,  et~al., 2020, \mn@doi [\ssr] {10.1007/s11214-020-00662-1}, 216,
  44

\bibitem[\protect\citeauthoryear{{El-Maarry} et~al.,}{{El-Maarry}
  et~al.}{2019}]{El-Maarry2019}
{El-Maarry} M.~R.,  et~al., 2019, \mn@doi [\ssr] {10.1007/s11214-019-0602-1},
  \href {https://ui.adsabs.harvard.edu/abs/2019SSRv..215...36E} {215, 36}

\bibitem[\protect\citeauthoryear{Filacchione et~al.,}{Filacchione
  et~al.}{2019}]{Filacchione2019}
Filacchione G.,  et~al., 2019, \mn@doi [\ssr] {10.1007/s11214-019-0580-3}, 215,
  19

\bibitem[\protect\citeauthoryear{{Giorgini} et~al.,}{{Giorgini}
  et~al.}{1996}]{Giorgini1996}
{Giorgini} J.~D.,  et~al., 1996, BAAS, \href
  {https://ui.adsabs.harvard.edu/abs/1996DPS....28.2504G} {28, 1158}

\bibitem[\protect\citeauthoryear{Grün et~al.,}{Grün et~al.}{2016}]{Grun2016}
Grün E.,  et~al., 2016, \mn@doi [\mnras] {10.1093/mnras/stw2088}, 462, S220

\bibitem[\protect\citeauthoryear{{Jehin} et~al.}{{Jehin}
  et~al.}{2011}]{Jehin2011}
{Jehin} E.,  et~al., 2011, The Messenger, \href
  {https://ui.adsabs.harvard.edu/abs/2011Msngr.145....2J} {145, 2}

\bibitem[\protect\citeauthoryear{{Jester} et~al.,}{{Jester}
  et~al.}{2005}]{Jester2005}
{Jester} S.,  et~al., 2005, \mn@doi [\aj] {10.1086/432466}, \href
  {https://ui.adsabs.harvard.edu/abs/2005AJ....130..873J} {130, 873}

\bibitem[\protect\citeauthoryear{{Jewitt} \& {Meech}}{{Jewitt} \&
  {Meech}}{1986}]{Jewitt1986}
{Jewitt} D.,  {Meech} K.~J.,  1986, \mn@doi [\apj] {10.1086/164745}, \href
  {https://ui.adsabs.harvard.edu/abs/1986ApJ...310..937J} {310, 937}

\bibitem[\protect\citeauthoryear{Jorda et~al.,}{Jorda et~al.}{2016}]{Jorda2016}
Jorda L.,  et~al., 2016, \mn@doi [Icarus]
  {https://doi.org/10.1016/j.icarus.2016.05.002}, 277, 257

\bibitem[\protect\citeauthoryear{Kelley \& Lister}{Kelley \&
  Lister}{2019}]{Kelley2019}
Kelley M.,  Lister T.,  2019, mkelley/calviacat,
  \mn@doi{10.5281/zenodo.2635840}, \url
  {https://doi.org/10.5281/zenodo.2635840}

\bibitem[\protect\citeauthoryear{Kelley et~al.,}{Kelley
  et~al.}{2021a}]{Kelley2021a}
Kelley M. S.~P.,  et~al., 2021a, \mn@doi [Planet. Sci. J.]
  {10.3847/psj/abfe11}, 2, 131

\bibitem[\protect\citeauthoryear{{Kelley} et~al.,}{{Kelley}
  et~al.}{2021b}]{Kelley2021b}
{Kelley} M. S.~P.,  et~al., 2021b, The Astronomer's Telegram, \href
  {https://ui.adsabs.harvard.edu/abs/2021ATel15053....1K} {15053, 1}

\bibitem[\protect\citeauthoryear{Knight, Snodgrass, Vincent, Conn, Skiff,
  Schleicher  \& Lister}{Knight et~al.}{2017}]{Knight2017}
Knight M.~M.,  Snodgrass C.,  Vincent J.-B.,  Conn B.~C.,  Skiff B.~A.,
  Schleicher D.~G.,   Lister T.,  2017, \mn@doi [\mnras]
  {10.1093/mnras/stx2472}, 469, S661

\bibitem[\protect\citeauthoryear{{Knollenberg} et~al.,}{{Knollenberg}
  et~al.}{2016}]{Knollenberg2016}
{Knollenberg} J.,  et~al., 2016, \mn@doi [\aap] {10.1051/0004-6361/201527744},
  \href {https://ui.adsabs.harvard.edu/abs/2016A&A...596A..89K} {596, A89}

\bibitem[\protect\citeauthoryear{{Lang}, {Hogg}, {Mierle}, {Blanton}  \&
  {Roweis}}{{Lang} et~al.}{2010}]{Lang2010}
{Lang} D.,  {Hogg} D.~W.,  {Mierle} K.,  {Blanton} M.,   {Roweis} S.,  2010,
  \mn@doi [AJ] {10.1088/0004-6256/139/5/1782}, \href
  {https://ui.adsabs.harvard.edu/abs/2010AJ....139.1782L} {139, 1782}

\bibitem[\protect\citeauthoryear{Lister et~al.,}{Lister
  et~al.}{2022}]{Lister2022}
Lister T.,  et~al., 2022, \mn@doi [Planet. Sci. J.] {10.3847/psj/ac7a31}, 3,
  173

\bibitem[\protect\citeauthoryear{Marschall et~al.,}{Marschall
  et~al.}{2020}]{Marschall2020}
Marschall R.,  et~al., 2020, \mn@doi [\ssr] {10.1007/s11214-020-00744-0}, 216,
  130

\bibitem[\protect\citeauthoryear{Mottola, Attree, Jorda, Keller, Kokotanekova,
  Marshall  \& Skorov}{Mottola et~al.}{2020}]{Mottola2020}
Mottola S.,  Attree N.,  Jorda L.,  Keller H.~U.,  Kokotanekova R.,  Marshall
  D.,   Skorov Y.,  2020, \mn@doi [\ssr] {10.1007/s11214-019-0627-5}, 216, 2

\bibitem[\protect\citeauthoryear{Pajola et~al.,}{Pajola
  et~al.}{2017}]{Pajola2017}
Pajola M.,  et~al., 2017, \mn@doi [Nat. Astron.] {10.1038/s41550-017-0092}, 1,
  0092

\bibitem[\protect\citeauthoryear{Schleicher}{Schleicher}{2010}]{Schleicher2010}
Schleicher D.,  2010, Composite Dust Phase Function for Comets, \url
  {https://asteroid.lowell.edu/comet/dustphase.html}

\bibitem[\protect\citeauthoryear{{Sharma}, {Kelley}, {Joharle}, {Kumar},
  {Swain}, {Bhalerao}, {Anupama}  \& {Barway}}{{Sharma}
  et~al.}{2021}]{Sharma2021}
{Sharma} K.,  {Kelley} M. S.~P.,  {Joharle} S.,  {Kumar} H.,  {Swain} V.,
  {Bhalerao} V.,  {Anupama} G.~C.,   {Barway} S.,  2021, \mn@doi [Res. Notes of
  the AAS] {10.3847/2515-5172/ac3ee4}, \href
  {https://ui.adsabs.harvard.edu/abs/2021RNAAS...5..277S} {5, 277}

\bibitem[\protect\citeauthoryear{Snodgrass, Tubiana, Bramich, Meech, Boehnhardt
   \& Barrera}{Snodgrass et~al.}{2013}]{Snodgrass2013}
Snodgrass C.,  Tubiana C.,  Bramich D.~M.,  Meech K.,  Boehnhardt H.,   Barrera
  L.,  2013, \mn@doi [A\&A] {10.1051/0004-6361/201322020}, 557, A33

\bibitem[\protect\citeauthoryear{Snodgrass et~al.,}{Snodgrass
  et~al.}{2016a}]{Snodgrass2016b}
Snodgrass C.,  et~al., 2016a, \mn@doi [\mnras] {10.1093/mnras/stw2300}, 462,
  S138

\bibitem[\protect\citeauthoryear{{Snodgrass} et~al.,}{{Snodgrass}
  et~al.}{2016b}]{Snodgrass2016a}
{Snodgrass} C.,  et~al., 2016b, \mn@doi [\aap] {10.1051/0004-6361/201527834},
  588, A80

\bibitem[\protect\citeauthoryear{Snodgrass et~al.,}{Snodgrass
  et~al.}{2017}]{Snodgrass2017}
Snodgrass C.,  et~al., 2017, \mn@doi [Phil. Trans. R. Soc. A]
  {10.1098/rsta.2016.0249}, 375, 20160249

\bibitem[\protect\citeauthoryear{Tonry et~al.,}{Tonry et~al.}{2012}]{Tonry2012}
Tonry J.~L.,  et~al., 2012, \mn@doi [ApJ] {10.1088/0004-637x/750/2/99}, 750, 99

\bibitem[\protect\citeauthoryear{{Tubiana} et~al.,}{{Tubiana}
  et~al.}{2015}]{Tubiana2015}
{Tubiana} C.,  et~al., 2015, \mn@doi [\aap] {10.1051/0004-6361/201424735}, 573,
  A62

\bibitem[\protect\citeauthoryear{Vincent et~al.,}{Vincent
  et~al.}{2016}]{Vincent2016}
Vincent J.-B.,  et~al., 2016, \mn@doi [\mnras] {10.1093/mnras/stw2409}, 462,
  S184

\bibitem[\protect\citeauthoryear{Vincent, Farnham, K{\"u}hrt, Skorov,
  Marschall, Oklay, El-Maarry  \& Keller}{Vincent et~al.}{2019}]{Vincent2019}
Vincent J.-B.,  Farnham T.,  K{\"u}hrt E.,  Skorov Y.,  Marschall R.,  Oklay
  N.,  El-Maarry M.~R.,   Keller H.~U.,  2019, \mn@doi [\ssr]
  {10.1007/s11214-019-0596-8}, 215, 30

\bibitem[\protect\citeauthoryear{{Zaprudin}, {Lehto}, {Nilsson}, {Pursimo},
  {Somero}, {Snodgrass}  \& {Schulz}}{{Zaprudin} et~al.}{2015}]{Zaprudin2015}
{Zaprudin} B.,  {Lehto} H.~J.,  {Nilsson} K.,  {Pursimo} T.,  {Somero} A.,
  {Snodgrass} C.,   {Schulz} R.,  2015, \mn@doi [\aap]
  {10.1051/0004-6361/201525703}, \href
  {https://ui.adsabs.harvard.edu/abs/2015A&A...583A..10Z} {583, A10}

\bibitem[\protect\citeauthoryear{{Zaprudin}, {Lehto}, {Nilsson}, {Somero},
  {Pursimo}, {Snodgrass}  \& {Schulz}}{{Zaprudin} et~al.}{2017}]{Zaprudin2017}
{Zaprudin} B.,  {Lehto} H.~J.,  {Nilsson} K.,  {Somero} A.,  {Pursimo} T.,
  {Snodgrass} C.,   {Schulz} R.,  2017, \mn@doi [\aap]
  {10.1051/0004-6361/201730475}, \href
  {https://ui.adsabs.harvard.edu/abs/2017A&A...604A...3Z} {604, A3}

\makeatother
\end{thebibliography}




\appendix

\section{Photometry}

\amend{This appendix gives all the photometry of 67P mentioned in this paper from both the 2015--2016 and 2021--2022 apparitions. All values are based on an aperture with radius $\rho$ = 10,000 km at the distance of the comet. Table~\ref{tab:photometry} lists all the data output from the pipeline processing, including the data that failed to run or was later discarded. It includes the calibrated $\textit{g-, r-, i-}$ and $\textit{z-}$band photometry for each frame of each telescope/instrument. It also includes the parameters zero point and colour correction term used in the magnitude calibrations. The geometric circumstances of the observations are also detailed.} 

\amend{Table~\ref{tab:nightly_average} gives the average of the $\textit{r-}$band photometry per night.}

\amend{Table~\ref{tab:LT2022} shows the $\textit{r-}$band photometry acquired at the LT during the 2021--2022 apparition. It includes the same parameters as Table~\ref{tab:photometry}. All magnitudes are calibrated to the PS1 $\textit{r}$-band from SDSS-$\textit{r}$ with a colour correction coefficient of -0.02 and assuming a $\textit{g-r}$ = 0.61 colour of the comet.}

\begin{table*}
    \centering
    \caption{$\textit{g-, r-, i-}$ and $\textit{z-}$band photometry. Measured within an aperture with radius $\rho$ 10,000 km. Full table is available online, first five rows given as an example.}
    \label{tab:photometry}
    \begin{tabular}{lllllllllllllll}
        \hline
        Date & MJD & Tel./Inst. & Filter & $m_{\textit{r}}$ & $\sigma_{m_{\textit{r}}}$ & $m_{\textit{g}}$ & $\sigma_{m_{\textit{g}}}$ & $m_{\textit{i}}$ & $\sigma_{m_{\textit{i}}}$   \\
        (UT) & (d) &  &  & (mag) & (mag) & (mag) & (mag) & (mag) & (mag) \\
        \hline
        2014-03-12T06:02:58.670 & 56728.2521 & NOT/ALFOSC & R\textunderscore Bes 650\textunderscore130 & 19.115 & 0.088 & -- & -- & -- & -- &     \\     
        2014-03-12T06:03:43.312 & 56728.2526 & NOT/ALFOSC & R\textunderscore Bes 650\textunderscore130 & 19.169 & 0.026 & -- & -- & -- & -- &   \\
        2014-03-12T06:09:12.809 & 56728.2564 & NOT/ALFOSC & R\textunderscore Bes 650\textunderscore130 & 19.134 & 0.026 & -- & -- & -- & -- &    \\
        2014-03-12T06:12:57.846 & 56728.2590 & NOT/ALFOSC & R\textunderscore Bes 650\textunderscore130 & 19.160 & 0.031 & -- & -- & -- & -- &    \\
        2014-03-12T06:16:43.103 & 56728.2616 & NOT/ALFOSC & R\textunderscore Bes 650\textunderscore130 & 19.312 & 0.053 & -- & -- & -- & -- &    \\
        ... & ... & ... & ... & ... & ... & ... & ... & ... & ...   \\ 
        \hline
    \end{tabular}
    \begin{tabular}{lllllllllllll}
    \hline
        $m_{\textit{z}}$ & $\sigma_{m_{\textit{z}}}$ & $m_{\textrm{Inst.}}$ & $\sigma_{m_{\textrm{Inst.}}}$ & ZP & $\sigma_{\textrm{ZP}}$ & C & Colour & RA & DEC & RA JPL & DEC JPL & Offset  \\
        (mag) & (mag) & (mag) & (mag)  & (mag) & (mag) &  & & (deg) & (deg) & (deg) & (deg) & (deg) \\
        \hline
        -- & -- & -10.115 & 0.087 & 29.145 & 0.014 & 0.14 & g-r & 291.11738 & -26.82187 & 291.11735 & -26.82069 & 0.00118 \\    
        -- & -- & -12.888 & 0.023 & 31.972 & 0.014 & 0.14 & g-r &  291.11737 & -26.82191 & 291.11746 & -26.82069 & 0.00122 \\
        -- & -- & -12.934 & 0.022 & 31.984 & 0.015 & 0.14 & g-r &  291.11735 & -26.82190 & 291.11821 & -26.82067 & 0.00145 \\
        -- & -- & -12.881 & 0.023 & 31.957 & 0.021 & 0.14 & g-r & 291.11735 & -26.82190 & 291.11873 & -26.82065 & 0.00175 \\
        -- & -- & -12.697 & 0.029 & 31.924 & 0.045 & 0.14 & g-r & 291.11734 & -26.82186 & 291.11925 & -26.82064 & 0.00209 \\
        ...& ...& ...& ...& ... & ... & ... & ... & ... & ... & ... & ... & ...   \\ 
        \hline
    \end{tabular}
    \begin{tabular}{lllllllll}
    \hline
        $\rho$  & $\rho$ & Pixel Scale & $r$ & $\Delta$ & Phase & Airmass & Seeing & Remarks \\
        (arcsec) & (pix) & (arcsec/pix) & (au) & (au) & (deg) &  & (arcsec) & \\
        \hline
        2.93 & 16.29 & 0.18 & 4.33 & 4.70 & 11.75 & 3.30 & -- & comet misidentified \\ 
        2.93 & 16.29 & 0.18 & 4.33 & 4.70 & 11.75 & 3.28 & -- & comet misidentified \\
        2.93 & 16.29 & 0.18 & 4.33 & 4.70 & 11.75 & 3.14 & -- & comet misidentified \\
        2.93 & 16.29 & 0.18 & 4.33 & 4.70 & 11.75 & 3.05 & -- & comet misidentified \\
        2.93 & 16.29 & 0.18 & 4.33 & 4.70 & 11.75 & 2.97 & -- & comet misidentified \\
        ...& ...& ... & ... & ... & ... & ... & ... & ... \\
        \hline
    \end{tabular}
\end{table*}

\begin{table*}
    \centering
    \caption{$\textit{r-}$band photometry averaged per night. Measured within an aperture with radius $\rho$ 10,000 km. Full table is available online, first five rows given as an example.}
    \label{tab:nightly_average}
    \begin{tabular}{lllllllllll}
        \hline
        Date & MJD & Telescope/Instrument & $m_{\textit{r}}$ & $\sigma_{m_{\textit{r}}}$ & $\rho$  & $\rho$  &  $r$ & $\Delta$ & Phase & Airmass\\
        (UT) & (d) &  &   (mag) & (mag) & (arcsec) & (pix) & (au) & (au) & (deg) &   \\
        \hline
        2015-08-14 & 57248 & CA 2.2-m/CAFOS & 13.444 & 0.113 & 7.78 & 14.97 & 1.2433 & 1.7709 & 33.89 & 1.99 \\
        2015-08-15 & 57249 & CA 2.2-m/CAFOS & 13.351 & 0.112 & 7.79 & 14.97 & 1.2435 & 1.7703 & 33.91 & 2.59 \\
        2015-08-16 & 57250 & CA 2.2-m/CAFOS & 13.452 & 0.026 & 7.79 & 14.98 & 1.2438 & 1.7697 & 33.92 & 2.57 \\
        2015-08-17 & 57251 & CA 2.2-m/CAFOS & 13.351 & 0.018 & 7.79 & 14.98 & 1.2443 & 1.7692 & 33.94 & 2.49 \\
        2015-08-18 & 57252 & CA 2.2-m/CAFOS & 13.382 & 0.028 & 7.79 & 14.98 & 1.2448 & 1.7688 & 33.95 & 2.60 \\
        ... & ... & ... & ... & ... & ... & ... & ... & ... & ... & ... \\ 
        \hline
    \end{tabular}
\end{table*}

\begin{table*}
    \centering
    \caption{$\textit{r-}$band photometry acquired at the LT between 2021--2022. Measured within an aperture with radius $\rho$ 10,000 km. Full table is available online, first five rows given as an example.}
    \label{tab:LT2022}
    \begin{tabular}{llllllllll}
        \hline
        Date & MJD & $m_{\textit{r}}$ & $\sigma_{m_{\textit{r}}}$ & $m_{\textrm{Inst.}}$ & $\sigma_{m_{\textrm{Inst.}}}$ & ZP & $\sigma_{\textrm{ZP}}$ & RA & DEC  \\ 
        (UT) & (d) & (mag) & (mag) & (mag) & (mag) & (mag) & (mag) & (deg) & (deg)  \\
        \hline
        2021-07-06T03:58:45.120 & 59401.1658 & 15.264 & 0.059 & -12.212 & 0.010 & 27.476 & 0.058 &  10.05779 & -0.32235 \\                              
        2021-07-06T03:59:28.320 & 59401.1663 & 15.279 & 0.063 & -12.195 & 0.010 & 27.474 & 0.063 &  10.05802 & -0.32229  \\                               
        2021-07-06T04:00:02.880 & 59401.1667 & 15.260 & 0.109 & -12.222 & 0.010 & 27.482 & 0.108 & 10.05823 & -0.32218   \\                               
        2021-07-06T04:00:46.080 & 59401.1672 & 15.242 & 0.120 & -12.215 & 0.010 & 27.457 & 0.119 & 10.05844 & -0.32208   \\                               
        2021-07-06T04:01:20.640 & 59401.1676 & 15.273 & 0.053 & -12.207 & 0.010 & 27.480 & 0.052 & 10.05869 & -0.32201  \\ 
        ... & ... &  ... & ... & ... & ... & ... & ... & ... & ...  \\ 
        \hline
    \end{tabular}
    \begin{tabular}{llllllllllll}
        \hline
        RA JPL & DEC JPL & Offset & $\rho$  & $\rho$  & Pixel Scale & $r$ & $\Delta$ & Phase & Airmass & Seeing & Remarks\\
        (deg) & (deg) & (deg) & (arcsec) & (pix) & (arcsec/pix) & (au) & (au) & (deg) &  & (arcsec) & \\
        \hline
        10.05841 & -0.32240 & 0.0006 & 9.46 & 31.55 & 0.30 & 1.85 & 1.46 & 33.29 & 1.59 & 1.45 & --\\
        10.05863 & -0.32231 & 0.0006 & 9.46 & 31.55 & 0.30 & 1.85 & 1.46 & 33.29 & 1.59 & 1.60 & --\\
        10.05886 & -0.32222 & 0.0006 & 9.46 & 31.55 & 0.30 & 1.85 & 1.46 & 33.29 & 1.58 & 1.54 & --\\
        10.05908 & -0.32213 & 0.0006 & 9.46 & 31.55 & 0.30 & 1.85 & 1.46 & 33.29 & 1.58 & 1.45 & --\\
        10.05930 & -0.32204 & 0.0006 & 9.46 & 31.55 & 0.30 & 1.85 & 1.46 & 33.29 & 1.58 & 1.54 & --\\
        ... & ... & ... & ... & ... & ... & ... & ... & ... & ... & ... & ... \\
        \hline
    \end{tabular}
\end{table*}



\bsp	
\label{lastpage}
\end{document}